# Electromagnetic Interaction in the Presence of Isotopic Field-Charges and a Kinetic Field

György Darvas

Symmetrion, Budapest, H-1067, 29 Eötvös St.
http://symmetry.hu/symmetrion/, darvasg@iif.hu

**Abstract** This paper is a continuation of the article "The Isotopic Field-Charge Assumption Applied to the Electromagnetic Interaction". It continues the discussion and consequences of the extended Dirac equation in the presence of isotopic mass and electric charges, and a kinetic gauge field. In compliance with the author's previous papers (Darvas 2009, IJTP 2011, IJTP 2013), there appears a second conserved Noether current in the interaction between two electric charges in the presence of isotopic electric charges and a kinetic field. This second conserved current involves the conservation of the isotopic electric charge spin, and that predicts the existence of quanta of the kinetic field (dions associated with the photons). It is concluded that with the discussed conditions, the electromagnetic interaction should be mediated by photons and their dion partners together. The conclusions give physical meaning, among others, to the electric moment and to a virtual coupling spin.

## 1 Introduction

The papers [8, 10] developed a general field-theoretical model for the conservation of the isotopic field-charge spin, applicable to different kinds of interactions. That model was based on the qualitative distinction between two types of the source charges (e.g., gravitational and inertial masses, etc.) of any physical field, and if so, it assumes interaction between the isotopes of the two types of field-charges. In 2011, the author started to *specify the model to individual interaction types*. Certain elements were predicted in a phenomenological way earlier [6, 7]. [17] started to specify the isotopic field-charge assumption to the electromagnetic interaction, and derived the extended Dirac equation in the presence of isotopic field-charges and a kinetic gauge field. This paper discusses further consequences to be derived from the so extended Dirac equation.

A few further papers by the author showed that the presence of a kinetic field with a velocity dependent metric and isotopic field-charges involve a (velocity arrowed) direction-dependent, anisotropic, that means, Finsler geometry [9-11, 13-16]. Gravity [35] as a Finslerian phenomenon has been discussed recently also by other authors (e.g., [2, 21]; see also appendices A and B in van 't Hooft's [36] without explicitly mentioning Finsler geometry).

The discussion in this paper is based on the determination of the electromagnetic field tensor adapted to the above conditions. We determine the *Hamiltonian* and the *Lagrangian* of the interacting two electrically charged particles in order to specify the *full momentums* in the presence of a kinetic field (sections 2 - 4). We define the *electromagnetic field tensor*, and the *curvature* (sections 5 - 6), as well as the *Lorentz force* under the above conditions (section 7). The *conserved currents* and the *conserved isotopic electric charge spin* are discussed in section 8, and we *quantise the kinetic field* (section 9). In the closing section (9.2), we shortly demonstrate how one can find the *mediating bosons* (the $\delta$ dions) of the kinetic field.

This paper extends the field theoretic model of the electromagnetic interaction started in [17] (cf., footnote 1 in [17]). To remember, [17] introduced the isotopic electric charges in the classical Maxwell EM theory and showed that – applied in itself alone – the presence of isotopic electric charges would destroy the Lorentz invariance of the Maxwell theory. To restore the broken invariance, we applied the conservation of the isotopic field-charge spin, proven in [10], so that the two transformations executed together ensure the invariance and save the physical relevance of our theory. Then we introduced the isotopic electric charges in the classical Dirac equation extended it with a kinetic field, what was introduced first in [8]. We determine the Hamiltonian and the Lagrangian demanded by the extended Dirac equation, the extended, full magneto-kinetic and electro-kinetic moments, their commutation rules, the extended electro-magnetic field tensor, the curvature of the connection field, the relevant conservation law, and then we quantise the kinetic field and predict the mediating boson of the derived kinetic interaction.

## 2 The Hamiltonian and the Lagrangian of the electromagnetic interaction in the presence of isotopic gravitational and electric charges as well as a kinetic gauge field

The paper [17] derived a modified Dirac equation

$$\begin{aligned} \{[-(p_4 + \frac{\rho_T}{c}A_4 + \frac{\rho_V}{c}D_4)^2 + (\mathbf{p} + \frac{\rho_V}{c}\mathbf{A} + \frac{\rho_V}{c}\mathbf{D})^2 + m_V^2 c^2] + \\ + \hbar(\boldsymbol{\sigma}, \text{rot}\left(\frac{\rho_V}{c}\mathbf{A}\right)) + i\hbar\gamma_5(\boldsymbol{\sigma}, \text{grad}\left(\frac{\rho_T}{c}A_4\right) + \frac{1}{c}\frac{\partial}{\partial t}\left(\frac{\rho_V}{c}\mathbf{A}\right)) + \\ + \hbar(\boldsymbol{\sigma}, \text{rot}\left(\frac{\rho_V}{c}\mathbf{D}\right)) + \hbar(\boldsymbol{\sigma}, \frac{\rho_V^2}{c^2}[D_j D_k - D_k D_j]) + \\ + i\hbar\gamma_5(\boldsymbol{\sigma}, \text{grad}\left(\frac{\rho_V}{c}D_4\right) + \frac{1}{c}\frac{\partial}{\partial t}\left(\frac{\rho_V}{c}\mathbf{D}\right)) + \gamma_5 \frac{\rho_V^2}{c^2}(\boldsymbol{\sigma}, D_4\mathbf{D} - \mathbf{D}D_4) + \\ + \gamma_4\left[-(p_4 + \frac{\rho_T}{c}A_4 + \frac{\rho_V}{c}D_4) + \gamma_5(\boldsymbol{\sigma}, \mathbf{p} + \frac{\rho_V}{c}\mathbf{A} + \frac{\rho_V}{c}\mathbf{D}) + \gamma_4 m_V c\right](m_T - m_V)c\} \psi = 0 \end{aligned} \qquad (1)$$

where $m_T$ and $m_V$, as well as $\rho_T$ and $\rho_V$ are the kinetic and potential masses and electric charge densities, respectively, and **D** denotes a velocity dependent field.

Line 5 of the Eq. (1) yields the Schrödinger wave equation $i\hbar\frac{\partial}{\partial t}\psi = \mathbf{H}\psi$, similar to the clue we followed in section 3 of [17]:

$$i\hbar\frac{\partial}{\partial t}\psi = \left[-\rho_T A_4 - \rho_V D_4 - \gamma_5(\boldsymbol{\sigma}, i\hbar c\, \text{grad} - \rho_V \mathbf{A} - \rho_V \mathbf{D}) + \gamma_4 m_V c^2\right]\psi \qquad (2)$$

and hence the Hamiltonian is

$$\mathbf{H} = -\rho_T A_4 - \rho_V D_4 - \gamma_5(\boldsymbol{\sigma}, i\hbar c\, \text{grad} - \rho_V \mathbf{A} - \rho_V \mathbf{D}) + \gamma_4 m_V c^2 \ .$$

The Lagrangian of the interaction field can be constructed from the Hamiltonian. So

$$\mathbf{L} = \rho_T A_4 + \rho_V D_4 - \gamma_5(\boldsymbol{\sigma}, i\hbar c\, \text{grad} - \rho_V \mathbf{A} - \rho_V \mathbf{D}) + \gamma_4 m_V c^2$$

Obviously, this expression differs from the classical one in the two terms. The latter include first the three-component **D**, and the fourth component of the vierbein, i.e., $D_4$, on the one hand, and the two isotopic electric charges, on the other hand.

As we showed in sec. 3 of [17], the condition for obtaining the Schrödinger equation was, that $m_T \gg m_V$, that means, an extreme relativistic situation. If $m_T \neq m_V$, one can divide the full Eq. (1) by $(m_T - m_V)$. One could obtain Eq. (2) in this way. The division by $(m_T - m_V)$ may cause an increase in the energy of the system when the difference between $m_T - m_V$ approaches to 0, unless the change in **D** does not counterbalance it. This means, first, that **D** must be a monotone function of the velocity, at second we can determine a limit of its monotone increase with the increase of the velocity.

## 3 Calculation of the full magnetic and electric moments

Similar to [17], let us write again Eq. (1) in the form of

$$[W+W^A+W^D-H(m_T-m_V)c]\psi=0 \tag{3}$$

Close to the rest, division of Eq. (1) or (3) by $(m_T - m_V)$ makes $W$ and $W^A$ high. This operation gets sense only at far relativistic velocities. Nevertheless, just in the case of low velocities, the role of $H(m_T - m_V)c$ can be neglected, since $(m_T - m_V) \to 0$. What is interesting for us, it is the role of $W^D$. Let's divide $W^D$ (4)

$$W^D = \hbar(\boldsymbol{\sigma}, \text{rot}\left(\frac{\rho_V}{c}\mathbf{D}\right)) + \hbar(\boldsymbol{\sigma}, \frac{\rho_V^2}{c^2}[D_jD_k - D_kD_j]) + \\
+ i\hbar\gamma_5(\boldsymbol{\sigma}, \text{grad}\left(\frac{\rho_V}{c}D_4\right) + \frac{1}{c}\frac{\partial}{\partial t}\left(\frac{\rho_V}{c}\mathbf{D}\right)) + \gamma_5 \frac{\rho_V^2}{c^2}(\boldsymbol{\sigma}, D_4\mathbf{D} - \mathbf{D}D_4) \tag{4}$$

by $(m_T - m_V)$:

$$(\frac{\hbar}{c}\frac{\boldsymbol{\sigma}}{m_T - m_V}, \rho_V \text{rot}\mathbf{D}) + (\frac{\hbar}{c}\frac{\boldsymbol{\sigma}}{m_T - m_V}, igC^i_{jk}\frac{\rho_V^2}{c}D_jD_k) + \\
+ i\gamma_5\left[\left(\frac{\hbar}{c}\frac{\boldsymbol{\sigma}}{m_T - m_V}, \rho_V \text{grad}\, D_4 + \frac{\rho_V}{c}\frac{\partial}{\partial t}\mathbf{D}\right) + \left(\frac{\hbar}{c}\frac{\boldsymbol{\sigma}}{m_T - m_V}, \frac{\rho_V^2}{\hbar c}(\mathbf{D}D_4 - D_4\mathbf{D})\right)\right]. \tag{5}$$

As we saw in [17], the last term can be disregarded, since according to our simplifying assumption $D_i$ and $D_4$ commute with each other. So can one do with grad $D_4$ which is 0. We get:

$$\rho_V(\frac{\hbar}{c}\frac{\boldsymbol{\sigma}}{m_T - m_V}, \operatorname{rot}\mathbf{D} + igC^i_{jk}\frac{\rho_V}{c}D_jD_k) + i\gamma_5\rho_V\left(\frac{\hbar}{c}\frac{\boldsymbol{\sigma}}{m_T - m_V}, \frac{1}{c}\frac{\partial}{\partial t}\mathbf{D}\right) \qquad (6)$$

Note that there appear in (6) only the potential (Coulomb) charges, and the mass difference between the kinetic and potential states. The expression (6) can be written also in the form:

$$(\frac{\hbar}{c}\frac{\boldsymbol{\sigma}}{m_T - m_V}, \mathbf{M}^D) + i\gamma_5(\frac{\hbar}{c}\frac{\boldsymbol{\sigma}}{m_T - m_V}, \mathbf{N}^D) =$$
$$= \rho_V(\frac{\hbar}{c}\frac{\boldsymbol{\sigma}}{m_T - m_V}, \operatorname{rot}\mathbf{D} + igC^i_{jk}\frac{\rho_V}{c}D_jD_k) + i\gamma_5\rho_V\left(\frac{\hbar}{c}\frac{\boldsymbol{\sigma}}{m_T - m_V}, \frac{1}{c}\frac{\partial}{\partial t}\mathbf{D}\right) \qquad (7)$$

where $\mathbf{M}^D$ and $\mathbf{N}^D$ are the same, as defined in Eq. (11c) in [17], considering the mentioned neglecting. The two terms in the left side of (7) are the additional *magneto-kinetic* and the additional *electro-kinetic* moments of the kinetic gauge field of the interaction. As we saw at the end of [17], in contrast to the classical Dirac theory, in the presence of a kinetic gauge field the electro-kinetic momentum cannot be disregarded. Added to the components which are calculated from the electromagnetic field, it may include real components, and in a properly-chosen reference frame it obtained physical meaning. This latter option was not considered in the classical QED (cf. footnote 5 in [17]).

The *full magnetic moment* of the interaction in the presence of the kinetic gauge field will be the sum of the magnetic moments that appeared in the QED and that additional one derived above:

$$(\frac{\hbar}{c}\frac{\boldsymbol{\sigma}}{m_T - m_V}, \mathbf{M}^{FULL}) = \rho_V(\frac{\hbar}{c}\frac{\boldsymbol{\sigma}}{m_T - m_V}, \operatorname{rot}(\mathbf{A} + \mathbf{D}) + igC^i_{jk}\frac{\rho_V}{c}D_jD_k) \qquad (8)$$

The f*ull electric moment*, similarly, can be written as:

$$i\gamma_5(\frac{\hbar}{c}\frac{\boldsymbol{\sigma}}{m_T - m_V}, \mathbf{N}^{FULL}) = i\gamma_5\left(\frac{\hbar}{c}\frac{\boldsymbol{\sigma}}{m_T - m_V}, \operatorname{grad}\rho_T A_4 + \frac{\rho_V}{c}\frac{\partial}{\partial t}(\mathbf{A} + \mathbf{D})\right) \qquad (9)$$

Note, that in $\mathbf{N}^{FULL}$, there appears also the kinetic charge density.

These $\mathbf{M}^{FULL}$ and $\mathbf{N}^{FULL}$ should commute with the Hamiltonian operator of the interacting two charges (cf. sec. 4 below and sec. 5.3 in [17]).

**4 The momentum in a kinetic field and the appearance of a virtual "coupling" spin**

Assuming a central field, where the vector potential is **A** = 0, but the kinetic field is non-zero, the Hamiltonian of the system of two interacting charges is the following:

$$\mathbf{F} = -(p_4 + \frac{\rho_V}{c} D_4) + \gamma_5(\boldsymbol{\sigma}, \mathbf{p} + \frac{\rho_V}{c} \mathbf{D}) + \gamma_4 m_V c.$$

This formula differs from the previous one in two terms: $-\frac{\rho_V}{c} D_4 + \gamma_5(\boldsymbol{\sigma}, \frac{\rho_V}{c} \mathbf{D})$.

Note the following two remarks! On the one hand, these two terms do not give an additional component to the commutation of **F** and $(m + \frac{1}{2}\hbar\sigma)$: $D_1$ and $D_3$ are 0; $\gamma_5\sigma_2 D_2 = -\sigma_2 D_4 = iD_4$, so the two terms annullate (neutralise) each other. On the other hand, while the coefficients of $i\hbar\gamma_5$ in (11d) in [17] annullate each other *by rule*, this latter annullation is *accidental* (cf., [40], [41]). This term is real, for $i\sigma_2$ is real. According to [6, 7] and further calculations, they give an additional

$$c^2 g \rho_V \frac{i}{4} \hbar \frac{\sigma_2}{m_T - m_V} \qquad (10)$$

spin to the term $(\mathbf{m} + \frac{1}{2}\hbar\boldsymbol{\sigma})$, where $g$ denotes the coupling constant of the electromagnetic interaction, and $m_T, m_V$ are the masses of the interacting electric charges. Since we consider *interaction* between two electric charges, *this virtual "coupling" spin appears only when there are at least two, interacting electric charges present*. They are present (virtually) only in bound states of two particles. They vanish in all other instances [7]. The arrows of the virtual "coupling" spins of two interacting electric charges are directed opposite (anti-parallel), so they compensate each other. Therefore, they are unobservable, and do not give any observable additional spin to the two interacting agents. Note also, that this term is becoming large close to the rest, and disappears at high energies. The latter means that the coupling force is much larger between two electric charges near to rest, than between those moving with relatively high velocities to each other.

The above formula (10) shows that it is the *rest charge*, and the *difference between the kinetic and potential* (rest) *masses* what play determining role in the strength of the interaction between two electrically charged particles.

**5 The field tensors of the EM and the kinetic fields**

*5.1 The field tensor of the EM field*

In accordance with [10], the obtained equations yield the classical QED fields in the absence of a kinetic **D** field. Thus the elements of the field tensor of the EM field, as well as the conserved current are of the same form, like we learned in our usual textbooks. This means, our results extend the results accepted in the Standard Model, but do not influence the validity of those equations in the absence of considering a kinetic field. They provide the same conserved quantities like we learned in the semi-classical theory, that means in this instance the electric charge. This conclusion coincides with all said in connection with $J^{(1)}$ in [10].

*5.2 The field tensor of the kinetic field*

The field tensor of the kinetic field can be obtained as:

$$F^{(2)\mu\nu}(x) = \frac{\partial D_{\dot\rho}\lambda_\mu^\rho}{\partial x_\nu} - \frac{\partial D_{\dot\sigma}\lambda_\nu^\sigma}{\partial x_\mu} + D_{\dot\rho}\lambda_\mu^\rho D_{\dot\sigma}\lambda_\nu^\sigma - D_{\dot\sigma}\lambda_\nu^\sigma D_{\dot\rho}\lambda_\mu^\rho, \qquad (11)$$

where $\lambda_\mu^\rho = \partial_\mu \dot x^\rho = \frac{\partial \dot x^\rho}{\partial x_\mu} = \dot x^\rho_{,\mu}$ (cf. Eq. (6) in [10]).

In similar like we obtained the elements of the field tensor for the EM field from the terms in the second line in (1) of [17], we can determine the elements of the kinetic field tensor from Eqs. (11)-(11b) of [17]. For this reason, we will use the expressions defined for $\mathbf{M}^D$ and $\mathbf{N}^D$ that denote the two components of the field strengths of the field's kinetic potential **D**. From

$$\mathbf{M}^D = \rho_V \operatorname{rot} \mathbf{D} + igC^i_{jk}\frac{\rho_V^2}{c}D_j D_k \quad \text{and} \quad \mathbf{N}^D = \rho_V \operatorname{grad} D_4 + \frac{\rho_V}{\hbar c}\frac{\partial}{\partial t}\mathbf{D} + \frac{\rho_V^2}{\hbar c}(\mathbf{D}D_4 - D_4\mathbf{D}) \quad (12)$$

one can construct the following tensor:

$$\rho_V F^{\mu\nu} = \begin{bmatrix} 0 & M_3^D & -M_2^D & -i\gamma_5 N_1^D \\ -M_3^D & 0 & M_1^D & -i\gamma_5 N_2^D \\ M_2^D & -M_1^D & 0 & -i\gamma_5 N_3^D \\ i\gamma_5 N_1^D & i\gamma_5 N_2^D & i\gamma_5 N_3^D & 0 \end{bmatrix}$$

where

$$\mathbf{M}_i^D = \partial_j \rho_V D_k - \partial_k \rho_V D_j + igC_{jk}^i \frac{\rho_V^2}{c} D_j D_k = \rho_V \left( \partial_j D_k - \partial_k D_j \right) + igC_{jk}^i \frac{\rho_V^2}{c} D_j D_k =$$

$$= \rho_V \text{rot}_i \mathbf{D}(\dot{x}) + igC_{jk}^i \frac{\rho_V^2}{c} D_j D_k = \rho_V \left[ (\partial_{\dot{\rho}} D_k) \lambda_j^{\dot{\rho}} - (\partial_{\dot{\rho}} D_j) \lambda_k^{\dot{\rho}} \right] + igC_{jk}^i \frac{\rho_V^2}{c} D_j D_k$$

(13)

and

$$\mathbf{N}_i^D = \partial_i \rho_V D_4 + \frac{1}{\hbar c} \partial_t \rho_V D_i + \frac{\rho_V^2}{\hbar c} D_i D_4 - \frac{\rho_V^2}{\hbar c} D_4 D_i = \rho_V (\partial_i D_4 + \frac{1}{\hbar c} \partial_t D_i) + \frac{\rho_V^2}{\hbar c} (D_i D_4 - D_4 D_i)$$

(14)

Considering that grad $D_4 = 0$ and $D_4$ commutes with $D_i$:

$$\mathbf{M}_i^D = \rho_V (\text{rot}_i \mathbf{D}(\dot{x}) + igC_{jk}^i \frac{\rho_V}{c} D_j D_k) = \rho_V \left[ (\partial_{\dot{\rho}} D_k) \lambda_j^{\dot{\rho}} - (\partial_{\dot{\rho}} D_j) \lambda_k^{\dot{\rho}} \right] + igC_{jk}^i \frac{\rho_V^2}{c} D_j D_k \quad (15)$$

and $\quad \mathbf{N}_i^D = \frac{\rho_V}{\hbar c} \partial_t D_i \quad$ (16)

## 6 The curvature of the connection field

The curvature of the connection field can be read from the coefficient of the covariant extension of the matrix terms in $F^{\mu\nu}$ (12)-(14). $\mathbf{M}^D_i$ in can be written also in the form (15), where the last two terms define a covariant commutation of the elements $D_i$:

$$\mathbf{M}_i^D = \rho_V (\text{rot}_i \mathbf{D}(\dot{x}) + ig\Gamma_{jk}^i D_j D_k).$$

Here $\Gamma_{jk}^i = C_{jk}^i \frac{\rho_V}{c}$ denotes that $\Gamma$ depends only on constants, while $D_i$ depend on the $\dot{x}^\mu$ four-velocity components. The latter $\dot{x}^\mu(x_\nu)$ corresponds to the functions marked by Dirac [20] as $y_\nu^\Lambda$ through which he defined the metric of the field.[1,2] The metric of the field is much simpler than we expected, while the velocity dependence is transferred to the components of the **D** field.[3]

---

[1] While the Dirac equation – introduced and discussed first in his 1928 and 1929 papers [18, 19] – is presented in almost all usual textbooks on QED and field theory, his extension published in 1962 is mentioned rarely (cf., [39]).

[2] [22] studies also the metric of Dirac's field theory with kinematic conditions, in a similar, but also a little bit different context.

## 7 The Lorentz force in the presence of a kinetic field

Similar to the classical EM model (cf., section 3 in [17]), in the presence of isotopic electric charges and a kinetic field the Lorentz force can be written in the following form:

$$F^\mu = F^{\mu\nu} \frac{1}{c} j_\nu = \frac{1}{\rho_V} \begin{bmatrix} 0 & M_3^D & -M_2^D & -i\gamma_5 N_1^D \\ -M_3^D & 0 & M_1^D & -i\gamma_5 N_2^D \\ M_2^D & -M_1^D & 0 & -i\gamma_5 N_3^D \\ i\gamma_5 N_1^D & i\gamma_5 N_2^D & i\gamma_5 N_3^D & 0 \end{bmatrix} \begin{bmatrix} \rho_T \frac{\dot{x}^1}{c} \\ \rho_T \frac{\dot{x}^2}{c} \\ \rho_T \frac{\dot{x}^3}{c} \\ i\rho_V \end{bmatrix} =$$

$$= \frac{1}{\rho_V} \begin{bmatrix} M_3^D \rho_T \frac{\dot{x}^2}{c} - M_2^D \rho_T \frac{\dot{x}^3}{c} + \gamma_5 N_1^D \rho_V \\ -M_3^D \rho_T \frac{\dot{x}^1}{c} + M_1^D \rho_T \frac{\dot{x}^3}{c} + \gamma_5 N_2^D \rho_V \\ M_2^D \rho_T \frac{\dot{x}^1}{c} - M_1^D \rho_T \frac{\dot{x}^2}{c} + \gamma_5 N_3^D \rho_V \\ i\gamma_5 N_1^D \rho_T \frac{\dot{x}^1}{c} + i\gamma_5 N_2^D \rho_T \frac{\dot{x}^2}{c} + i\gamma_5 N_3^D \rho_T \frac{\dot{x}^3}{c} \end{bmatrix} =$$

$$= \frac{1}{c\rho_V} \begin{bmatrix} M_3^D \dot{x}^2 - M_2^D \dot{x}^3 & c\gamma_5 N_1^D \\ -M_3^D \dot{x}^1 + M_1^D \dot{x}^3 & c\gamma_5 N_2^D \\ M_2^D \dot{x}^1 - M_D \dot{x}^2 & c\gamma_5 N_3^D \\ i\gamma_5 N_1^D \dot{x}^1 + i\gamma_5 N_2^D \dot{x}^2 + i\gamma_5 N_3^D \dot{x}^3 & 0 \end{bmatrix} \begin{bmatrix} \rho_T \\ \rho_V \end{bmatrix} = \frac{1}{c\rho_V} H^{D\kappa l} \rho_l$$

where ($\kappa = 1, .., 4$), ($l = 1, 2$), or in the form

$$F^\mu = \frac{1}{c} \begin{bmatrix} M_3^D \dot{x}^2 - M_2^D \dot{x}^3 & c\gamma_5 N_1^D \\ -M_3^D \dot{x}^1 + M_1^D \dot{x}^3 & c\gamma_5 N_2^D \\ M_2^D \dot{x}^1 - M_D \dot{x}^2 & c\gamma_5 N_3^D \\ i\gamma_5 N_1^D \dot{x}^1 + i\gamma_5 N_2^D \dot{x}^2 + i\gamma_5 N_3^D \dot{x}^3 & 0 \end{bmatrix} \begin{bmatrix} \rho_T \\ \rho_V \\ 1 \end{bmatrix}. \qquad (17)$$

---

[3] [30] noticed that there is not the presence of a scalar field which affects the geometry of the space-time, although it changes the matter distribution.

It is obvious in the latter form that the isotopic electric charges do not concern the electric moment. Their ratio plays the role of a coefficient to the magneto-kinetic moment only. This ratio depends only on the Lorentz transformation, in which there appears the relative velocity of the two interacting charges to each other. This expression for the Lorentz force shows that our $\Gamma$ curvature obtained for the kinetic field is in its form similar to the $\Gamma$ curvature for the EM field as determined by Landau and Lifshitz ([25] §85).

The kinetic addition to the Lorentz force can be defined with the use of the above $F^{\mu\nu}$:

$$F^\mu = \frac{1}{c\rho_V}\begin{bmatrix}\left[M_k^D \times \dot{x}^j\right] & c\gamma_5 N_i^D \\ i\gamma_5 N_i^D \dot{x}^i & 0\end{bmatrix}\begin{bmatrix}\rho_T \\ \rho_V\end{bmatrix} = \frac{1}{c\rho_V}H^{D\kappa l}\rho_l \qquad (18)$$

This expression for the Lorentz force indicates that the weak intermediate bosons can be derived from the [4 x 2] matrix in the first square bracket [ ]: the photon $\gamma$, with mass zero, is associated with $H^{D22}$, $W^\pm$ with $H^{D12}$ and $H^{D21}$, while $Z^0$ with $H^{D11}$. Please, note the asymmetry between $H^{D12}$ and $H^{D21}$, what confirms the assumption by C. Møller [27], and what was indicated by S. Weinberg [37] in another way. Note also that we indicated the unification of the *electromagnetic* and the *weak interactions* in a different way than Weinberg did. Nevertheless, this latter is the theme of another paper.

**8 The conserved currents and the conserved isotopic electric charge spin**

In the possession of the Lagrangian and the field tensor, with the help of the $D_i$ kinetic potential field components, one can apply the Eq. (5) in [10]

$$J_\alpha^{(2)\nu}(x) = ig\left[\frac{\partial L}{\partial(\partial_\mu A_{4k})}(T_\alpha)_{kl}A_{4l}(\dot{x})\lambda_\mu^\nu - C_{\alpha\beta}^\gamma D_{\dot{\omega}\beta}(\dot{x})\lambda_\mu^\omega \times F_\gamma^{(2)\mu\nu}(x)\right]. \qquad (19)$$

In the instance (of the electromagnetic interaction) α = 1, $(T_\alpha)_{kl}$ is a unit matrix, and $C^\gamma_{\alpha\beta}$ are the three structure constants for the kinetic field, which commute the generators of the SU(2) symmetry group. $J^{(2)\nu}$ are the components of the conserved isotopic electric charge spin current, which include the contribution of the kinetic **D** field (cf., the second term on the right side)[4] (**J**$^{(2)}$= **J**$^D$). We have introduced the **D** field – which is shown to be responsible for the isotopic field charge spin transformation – to counteract the dependence of a *V* transformation on $J^{D_{1,2,3}}$ [10]. The field equations, which are satisfied by the twelve independent components of the **D** field, and their interaction with any field that carries isotopic field charge spin, are unambiguously determined by the defined current and the covariant $F^{(2)\mu\nu}$-s constructed from the components of **D**. Considering a general Lorentz- and gauge invariant Lagrangian, we obtain from the equations of motion that $J^{D_{1,2,3}}$ and $J^{D_4}$ are, respectively, the isotopic field charge spin current density and isotopic electric charge spin ($\Delta$) density of the system. The total isotopic electric charge spin

$$\Delta = \frac{i}{g}\int J^{D_4} \mathrm{d}^3 x \qquad (20)$$

is independent of time and independent of Lorentz transformation. $J^{D\mu}$ does not transform as a vector, while $\Delta$ transforms as a vector under *rotations in the isotopic electric charge spin field*. The $\Delta_{el}$ quanta of the **D** field (for EM) can be determined by insertion in the expression $J^{D_4}$.

---

[4] Similar attempts (like our in the velocity space) were made by [28] in the phase space (with a particular mapping from the configuration space to phase space), and they anticipated the quantization of the models.

$$\Delta_{el} = -\int\left[\frac{\partial L}{\partial(\partial_\mu A_4)}A_4\partial_\mu \dot{x}^4 - \gamma_5 C^\gamma D_{\dot{\rho}\gamma}\partial_i \dot{x}^\rho \times N_i\right]\mathrm{d}^3 x \qquad (C^\gamma = 0, \pm 1;\ \gamma = 1, 2, 3,\ \mu = 1, ..., 4)$$

I showed in ([10], section 4.2) that beside the conserved electromagnetic current (which provides the conservation of the electric charge, and the $\gamma$ quantum of the EM field), there exists another conserved current, $J^{D_4}$, in the presence of a kinetic gauge field **D** (see also section 9).

**9 Quantisation**

The coupling of a conserved quantity in a space-time dependent field (which coincides with one of our known physical fields) with another (in a kinetic, i.e., velocity dependent [1]) gauge field indicates that the derived conservation verified the invariance between two isotopic states of the field charges, namely between the potential and the kinetic electric charges. Remember that the conserved isotopic electric charges belong to the electromagnetic field, while $\Delta$ represents a single quantity belonging to the kinetic gauge field **D.**

In short, in the presence of kinetic fields we have two conserved currents that are effective simultaneously [10]. The kinetic gauge field **D** is present simultaneously with the interacting matter- ($A_4$) and gauge- (**A**) fields.[5] The presence of **D** corresponds to the property of the electric field charges that they split in two isotopic states. And (analogously to the isotopic spin) these two states are given the name *isotopic electric charge spin* what we denoted by $\Delta_{el}$. The source of the isotopic electric charge spin ($\Delta_{el}$) is the field $D_4(\dot{x})$, in interaction with the kinetic gauge field **D**.

The physical meaning of $\Delta_{el}$ can be understood by the specification of the transformation group associated with the **D** field, which describes the transformations of the isotopic electric charges. They can take two (potential and kinetic) isotopic density states $\rho_V$ and $\rho_T$ in a simple unitary abstract space. Their symmetry group is SU(2), that can be represented by 2x2 $T_\alpha$ matrices [26]. There are three independent $T_\alpha$ that may transform into each other, following the rule $\left[T_\alpha, T_\beta\right] = iC_{\alpha\beta}^\gamma T_\gamma$, where the structure constants can take the values 0, ±1. Let $T_1$ and $T_2$ be those which do not commute with $T_3$; they generate transformations that mix the different values of $T_3$, while this "third" component's eigenvalues represent the members of a $\Delta_{el}$ doublet. For the isotopic field charges compose a charge density doublet of $\rho_V$ and $\rho_T$, the field's wave function can be written as

$$\psi = \begin{pmatrix} \psi_T \\ \psi_V \end{pmatrix}. \tag{21}$$

Expression (21) is the wave function for a single electron which may be in the "potential state", with amplitude $\psi_V$, or in the "kinetic state", with amplitude $\psi_T$.[6] $\psi$ in (21) represents a mixture of the potential and kinetic states of the electric charge, and there are $T_\alpha$ that govern the mixing of the components $\psi_V$ and $\psi_T$ in the transformation. $T_\alpha$ ($\alpha$ = 1, 2, 3) are representations of operators which can be taken as the three components of the isotopic electric charge spin, namely $\Delta_1$, $\Delta_2$, $\Delta_3$ that follow the same (non-Abelian) commutation rules as do the $T_\alpha$ matrices, [$\Delta_1$, $\Delta_2$] = $i\Delta_3$, etc. These operators represent the charges of the isotopic electric charge spin space, and $\psi$ are the fields on which the operators of the gauge fields act.

---

[5] Concerning the parallel presence of a scalar and a kinetic gauge field, authors of [31] enlarge the configuration space by including a scalar field additionally, and taking anisotropic models into account too. They also investigated whether Noether's symmetry holds in the known form alone or doesn't, that means, whether the single Noether current should be extended (let us add: with another current).

[6] Note, that the potential (Coulomb) charges behave like corpuscles, while the kinetic (Lorentz type) charges like waves [12]. This complementary double behaviour (formulated first by Bohr in 1927, then discussed in 1937 [46]) became subject of studies again (cf., [29]).

The quanta of the **D** field should carry isotopic electric charge spin $\Delta_{el}$. The $\Delta_{el}$ doublet, as a conserved quantity, is related to the two isotopic states of the electric charges, and the associated operators ($\Delta_i$) induce transitions from one member of the doublet to the other.

The *invariance between $\rho_V$ and $\rho_T$* (what is ensured by the conservation of $\Delta_{el}$), and their abilitiy to swap, *means* also that *they can restore the symmetry in the physical equations which appeared to get lost when we replaced the general $\rho$ by their isotopes $\rho_V$ and $\rho_T$.*

### 9.1 The predicted quanta (dion $\delta_{el}$)

Summarising the above results in short, when we introduced the distinguished two kinds of isotopic electric charges in our physical equations we reached a limit: we obtained equations, where certain symmetries of the traditional equations were distorted. That was not in accordance with our experience. Then, we derived a conservation law for the newly introduced quantity, the isotopic electric charge spin ($\Delta_{el}$), and its invariance (applied together with the Lorentz invariance) restored the lost symmetry. The isotopic electric charge spin should exist in an above defined gauge field **D**.

Such a gauge field must have quanta that carry the isotopic electric charge spin $\Delta_{el}$.[7] Exchange of these quanta should mediate between the interacting field charges, so that switch the emitter electric charge $\rho_V \rightarrow \rho_T$, and the recipient electric charge from $\rho_T \rightarrow \rho_V$ and *vice versa*. This holds, because according to field theories, any conserved property presumes the existence of a mediating boson. [10] demonstrated that in this case too, a mediating boson – called "dion" – must belong to each isotopic field-charge pair. (The 13 dions are the following: all the graviton, the photon, the two charged W and the neutral Z that mediate the weak interaction, and the eight gluons that mediate the strong interaction should have a dion brother.) The conservation of $\Delta_{el}$ involves the prediction of a dion associated with the electromagnetic field, a boson-brother of the photon. The theory [10] describes the necessity for the existence of these dions.

---

[7] Jackiw and Rebbi [23] (with assistance by t'Hooft) investigating the YM pseudoparticle solution by [3], which was found to be O(5) invariant, and after having applied the solution to the Dirac equation foresaw and demonstrated that the pseudoparticle was distinguished by possessing a large *kinematical invariance group* possibly important in future developments of the theory, although they did not analyse these kinetic consequences in particular for the Dirac equation.

We denoted the *predicted* quanta of the **D** field by *δ* [10].[8] The $δ_{el}$ quanta (dions) carry the *Δ*$_{el}$ (isotopic electric charge spin as a physical property: the charge of the **D** field). For all the quantum numbers of the interacting field charges are mediated by the respective SM mediating bosons, the quantum numbers of the dion (*δ*) are 0 but the *Δ* (and, of course, they carry mass). We can agree in a free convention for the sign of *Δ*$_{el}$ to be + or − (½) according to whether it switches the isotopic field charge spin from a potential state to a kinetic or back.

*9.2 Observation of the predicted quanta δ*

The observation of dions is possible in high energy collisions. The effect of the kinetic field is more apparent at high energies (high velocities). At higher energies the mass exchange between the two interacting isotopic field charges ($m_T - m_V$) is higher, due to the increasing relativistic difference between $m_T$ and $m_V$. The effective cross-section increases with the increase of the energy of the interaction, and the probability to find the searched dions increases with the energy.

Observation of the quantum $δ_{gr}$ for the gravitational field seems unplausible, since one has never seen any graviton. Observation of the electric $δ_{el}$ and the three weak dions is more probable, but the probability of the realisation is low, although not excluded. The Feynman diagram of a dion exchange for electric charges (see [10]) is the following:

---

[8] We called this hypothetical boson "*dion*", after the Greek term meaning 'flee', 'flight', 'rout' in English. The name dion brings about some associations of the *dyons*, which were proposed first by Schwinger, 1969 [32]. Dyon was presumed as a hypothetical particle endowed with both electric and magnetic charges. From our aspect it can be considered as the first idea of doubling the properties of a charged physical object. Although that doubling of properties differed from our distinction between isotopic electric charges that had there a similar feature: it distinguished the Coulomb-like charges from magnetic charges, which were assumed as results of the velocity of current-like (kinetic) charges. Schwinger introduced this concept when he extended the quantization condition, set up earlier by Dirac, to the dyon. By the dyon model, Schwinger predicted a particle with the properties of the J/ψ meson, years before it was discovered in 1974. A sign of the actuality of the topic is a renaissance of the discussion of dyons in the literature of non-Abelian theories [33] and [5]. The latter refers to octonion elecrodynamics, what is widely used in the literature, and what we reduced to quaternions in [8]. (Concerning earlier treatments, cf. e.g., [34, and 24].)

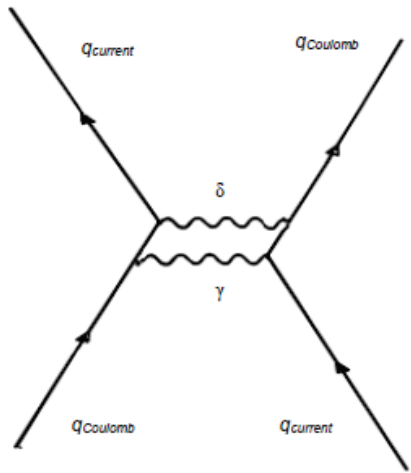

**Figure 1**: Feynman diagram of parallel photon and dion exchanges in an interacion between two electric charges

It seems more probable to demonstrate the existence of the eight strong dions, in an indirect way, but that would be the theme of another paper.

## 10 Conclusions

The present paper extends the results discussed in [17] with a few additional issues.

Based on the calculation of the extended, full magnetic and kinetic moments in an electromagnetic interaction, we could calculate a new Hamiltonian and Lagrangian of the interaction (section 2), and the full magneto-kinetic and electro-kinetic moments (section 3). Both are real, measurable physical quantities. This restored the physical meaning of the electric moment rejected by Dirac in [18] (cf., footnote 5 in [17].

The determination of the momentums of the kinetic field and application of the commutation rules provided an additional "virtual" coupling spin (Eq. (10)), which is unobservable, because it occurs only in interaction between two charges, never in single particles, and is anti-parallel at the two interacting agents' sides, that means, the two values compensate each other (section 4).

The appearance of this additional, "virtual" coupling spin represents an accidental symmetry (discussed in Weinberg [40], pp. 13-14). The existence of such an additional coupling spin was predicted in [7], but its exact value is derived first in this paper.

There is the *rest charge*, and the *difference between the kinetic and potential* (rest) *masses* that play determining role in the strength of the interaction between two electrically charged particles.

The extended Dirac equation led to derive additional terms to the Lorentz force. Eq. (18) includes a tensor $H^D$ whose elements allow us to conclude the weak intermediate bosons in a different way than Weinberg's original paper [37] did, including the massless photon and the three massive weak bosons.

The value of the isotopic electric charge spin is determined (section 8).

The invariance between the potential and kinetic charges (what is ensured by the conservation of the isotopic electric charge spin) and their abilitiy to swap means also that they can restore the symmetry in the physical equations which appeared to get lost when we replaced the general charge by their two isotopic states (section 9).

– . –

As S. Weinberg formulated at the end of his Nobel lecture [38] "... quantum field theory, which was born just fifty years ago from the marriage of quantum mechanics with relativity, is a beautiful but not very robust child. ... at superhigh energies is susceptible to all sorts of diseases ... and it needs special medicine to survive. One way that a quantum field theory can avoid these diseases is to be renormalisable and asymptotically free, but there are other possibilities. ... Thus, one way or another, I think that quantum field theory is going to go on being very stubborn, refusing to allow us to describe all but a small number of possible worlds, among which, we hope, is ours." The above paper (together with [17]) attempted "another" way to present a predicted "possibility".